\documentclass[prb,twocolumn,showpacs,superscriptaddress]{revtex4}
\usepackage{graphicx}
\usepackage{amsmath}
\usepackage{bm}
\usepackage{color}
\usepackage{hyperref}
\begin{document}
\title{Supersolids in one dimensional Bose Fermi mixtures}

\author{Fr\'ed\'eric H\'ebert}
\author{George G. \surname{Batrouni}}
\author{Xavier Roy}

\affiliation{INLN, Universit\'e de Nice-Sophia Antipolis, CNRS, 1361
  route des Lucioles, 06560 Valbonne, France}

\author{Val\'ery G. \surname{Rousseau}}
\affiliation{Instituut-Lorentz, LION, Universiteit Leiden, Postbus
  9504, 2300 RA Leiden, The Netherlands}

\date{\today}

\begin{abstract}
Using quantum Monte Carlo simulations, we study a mixture of bosons
and fermions loaded on an optical lattice. With simple on-site
repulsive interactions, this system can be driven into a solid
phase. We dope this phase and, in analogy with pure bosonic systems,
identify the conditions under which the bosons enter a supersolid
phase, \textit{i.e.}, exhibiting at the same time charge density wave
and superfluid order.  We perform finite size scaling analysis to
confirm the presence of a supersolid phase and discuss its properties,
showing that it is a collective phase that also involve phase
coherence of the fermions.
\end{abstract}

\pacs{67.80.Kb, 67.85.Pq, 05.30.-d}

\maketitle

The supersolid (SS) phase in which superfluid and solid order coexist
was first examined in $^4$He systems more than 50 years ago
\cite{penrose56}.  The presence of a SS phase was conclusively
demonstrated in several lattice models such as variants of the bosonic
Hubbard model.  However, this phase has not been unambiguously
observed experimentally and, with the recent progress in low
temperature physics, the search for experimental evidence of a SS
phase has been reinvigorated.

In solid helium, a non-classical reduction of the moment of inertia
was observed in torsional oscillator measurements.\cite{kim04,kim04-2}
This reduction is due to the appearance of a superfluid fraction in
the material and has been first interpreted as a sign of a SS phase.
However, several experimental \cite{rittner06,rittner07,sasaki06} and
numerical studies \cite{boninsegni06,boninsegni07,boninsegni07-2}
suggested that this superfluid behavior was in fact due to the
presence of non local defects in the system (grain boundaries for
examples) along which superfluid currents exist. The presence of a
bulk supersolid phase thus appears to be questionable in these
systems.

Another promising approach for finding a SS phase emerged in the
context of cold atoms loaded on optical lattices. Such systems, being
well described by bosonic Hubbard models, are a good starting point to
look for supersolids.  However the conditions necessary for
supersolids to appear in these models are not easily achieved in real
experiments: One generally observes a direct solid-superfluid
transition \cite{batrouni90} or a coexistence of superfluid and solid
(in the case of a first order phase transition)\cite{batrouni01}.
Supersolids are stabilized in these models by specific long range
interactions
\cite{sengupta05,melko05,wessel05,heidarian05,boninsegni05,batrouni06,pollet08}
or long range hopping terms.\cite{chen08} Engineering precisely the
values of the interaction as a function of distance appears to be
tricky to achieve.

Recently Bose-Fermi mixtures have been introduced as a way to study
the physics of fermions, using sympathetic cooling with the bosons to
reach very low temperatures.\cite{sympa1,sympa2} However several
theoretical studies
\cite{kuklov03,duan03,sengupta05-2,pollet06,mathey04,mathey07} have
shown that such mixtures loaded on optical lattices have a rich phase
diagram where collective phases of fermions and bosons
appear. Interestingly, with simple repulsive on-site interactions
between fermions and bosons, the system can be driven in a solid phase
where density-wave order develops.\cite{pollet06} We will study here
the doping of such a solid phase and see under what conditions it
could be driven into a bosonic supersolid phase. Other recent studies
suggested the presence of a SS behavior in Bose-Fermi
mixtures. \cite{titvinidze08,buchler03,mathey08,mathey07-2}

The paper is organised as follow. In section I, we introduce the model
and discuss the solid phase which is the starting point of our study.
In section II, we explore the different ways of doping this system and
determine which one could lead to a supersolid phase.  Finally, in
section III, we perform finite size scaling analysis of physical
quantities to verify if the supersolid phase persists in the
thermodynamic limit.

\section{Solid Phase in the Bose-Fermi Hubbard model}

We study a one dimensional Hubbard model for a mixture of bosons and
polarized (spinless) fermions. The Hamiltonian is given by:
\begin{eqnarray}
\mathcal{H} &=& \sum_{r=1}^L \left( -t_{\text{b}}\, b^\dagger_{r+1}
b_r -t_{\text{f}}\, f^\dagger_{r+1} f_r + \text{h.c.}
\right)\nonumber \\ && + \sum_{r=1}^L \left( U_{\text{bb}}\,
\frac{n^{\text{b}}_r \left(n^{\text{b}}_r-1\right)}{2} + U_{\text{bf}}\,
n^{\text{b}}_r n^{\text{f}}_r \right)
\label{hamiltonian}
\end{eqnarray}
where $b^\dagger_{r}$ ($b_{r}$) creates (destroys) a boson on site $r$
while $f^\dagger_{r}$ and $f_r$ are the corresponding operators for
fermions. The first term in Eq.(\ref{hamiltonian}) describes tunneling
of bosons and fermions between neighboring sites, with different
associated energies $t_{\text b}$ and $t_\text{f}$.  In the following,
the energy scale is fixed by choosing $t_\text{b}=1$.
$n^\text{b}_{r}$ and $n^{\text{f}}_r$ are the bosonic and fermionic
number operators at site $r$ and $U_{\text{bb}}$ and $U_{\text{bf}}$
are the boson-boson and boson-fermion contact repulsion terms.  $L$ is
the total number of sites in a one-dimensional chain.

We will be mostly interested in the behavior of bosons and will,
therefore, study the bosonic Green function $G_\text{b}(R)$ which
measures phase correlations:
\begin{equation}
G_\text{b}(R) = \frac{1}{L}\sum_{r=1}^L \left\langle b^\dagger_{r+R}\,
b_{r} + b^\dagger_{r} b_{r+R} \right\rangle
\end{equation}
as well as the boson superfluid density
$\rho_\text{s} = \langle W^2 L\rangle/ (2 \beta t_\text{b})$ where $W$
is the winding number\cite{pollock87}  and $\beta$ the inverse temperature.
The density-density correlation
function
\begin{equation}
D_\text{b}(R) = \frac{1}{L} \sum_{r=1}^L \left[ \left\langle
  n^{\text{b}}_{r+R} \, n^{\text{b}}_r \right\rangle - \left\langle
  n^{\text{b}}_{r+R} \right\rangle \left\langle n^{\text{b}}_r
  \right\rangle\right]
\end{equation}
and its Fourier transform, the structure factor $S_\text{b}(k)$, give
information on possible solid density-wave order.

We study this model using two different versions of the worm
algorithm: the canonical worm algorithm (CW)
\cite{rombouts06,vanhoucke06} and the directed stochastic Green
function algorithm (DSGF) \cite{rousseau08, rousseau08-2}.  The CW
method, which is very efficient for measurement of equal time Green
functions, has to be modified to include simultaneous updates of
bosons and fermions to allow the study of mixtures \cite{pollet05}.
However, this algorithm sometimes becomes inefficient for large system
sizes ($L>20$) especially failing to sample different winding numbers
which is necessary to calculate the superfluid density
\cite{pollock87}.  In those cases we used the DSFG which explores a
much larger configuration space and thus allows more efficient
fluctuations of the winding number and the measure of $\rho_\text{s}$.
The two algorithms gave results in agreement for other quantities in
the range of sizes we used.

\begin{figure}[h]
\includegraphics[width=8.5cm]{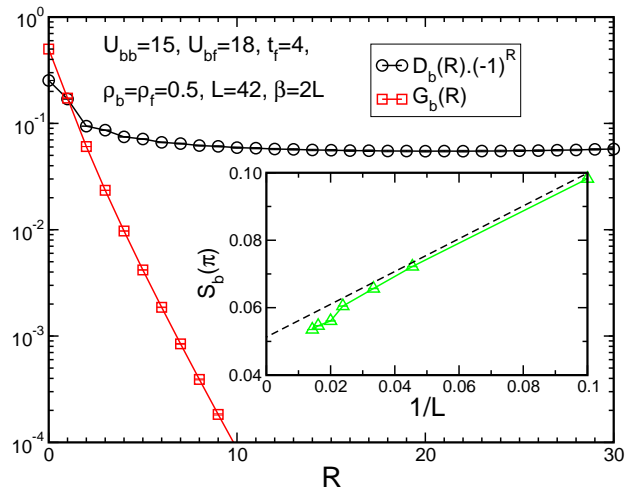}
\caption{(Color online) Density-density correlation $D_\text{b}(R)$
  (multiplied by $(-1)^R$)and Green function $G_\text{b}(R)$ for the
  bosons in the solid ordered phase. Inset: Finite size behavior of
  the bosonic structure factor $S_\text{b}(\pi)$ for the same
  parameters and sizes ranging from $L=10$ to 70.\label{solid}}
\end{figure}

This model was widely studied in the special case where the total
number of particles equals the number of sites $N_\text{b} +
N_\text{f} = L$ where $N_\text{b(f)}$ is the number of bosons
(fermions).\cite{pollet05,sengupta05-2} For large enough values of the
repulsions, double occupancy of sites is forbidden and one can
describe the system in terms of a pseudo spin-1/2 $\sigma_{r}^z =
n^\text{b}_r - n^\text{f}_r = \pm 1$. \cite{kuklov03,duan03} The
Hamiltonian can then be mapped, in the low energy limit, into an
effective Heisenberg Hamiltonian
\begin{equation}
\mathcal{H}_\text{eff} = 
\sum_r J_{xy} \left(\sigma^x_r \sigma^x_{r+1} + \sigma^y_r \sigma^y_{r+1} \right)+ \sum_r
J_z \sigma^z_r \sigma^z_{r+1}\label{pseudospin}
\end{equation}
where $J_{xy} = -t_\text{b} t_\text{f} / U_\text{bf} $ and $J_z =
(t_\text{b}^2+t_\text{f}^2)/(2U_\text{bf}) -
t^2_\text{b}/U_\text{bb}$.  When $J_z > |J_{xy}|$, the pseudo-spin
system enters a N\'eel antiferromagnetic phase along the $z$ axis.\cite{giam03} 
In terms of bosons, this antiferromagnetic order corresponds to a density
wave order with alternating occupied and empty sites (see
Fig.~\ref{solid}). The Green function, $G_\text{b}(R)$, decays
exponentially indicating the absence of phase coherence. This phase
persists in the thermodynamic limit at zero temperature (see
Fig.~\ref{solid}, inset).

\section{Doping the solid phase}

A similar density wave ordered phase for $N_\text{b} = L/2$ is
observed in a one-dimensional bosonic system with near-neighbour
repulsion.\cite{batrouni06} In this latter system a supersolid phase
can be present when the solid phase is doped by adding
bosons. However, for a supersolid phase to appear, the interactions
must be chosen so that the added bosons do not introduce defects in
the previous solid order: They must preferentially come on sites
already occupied by bosons. A mean field argument yields that the
repulsion between bosons located on neighboring sites must be greater
than half the on-site repulsion $U_\text{bb}$.

\begin{figure}[h]
\includegraphics[width=8.5cm]{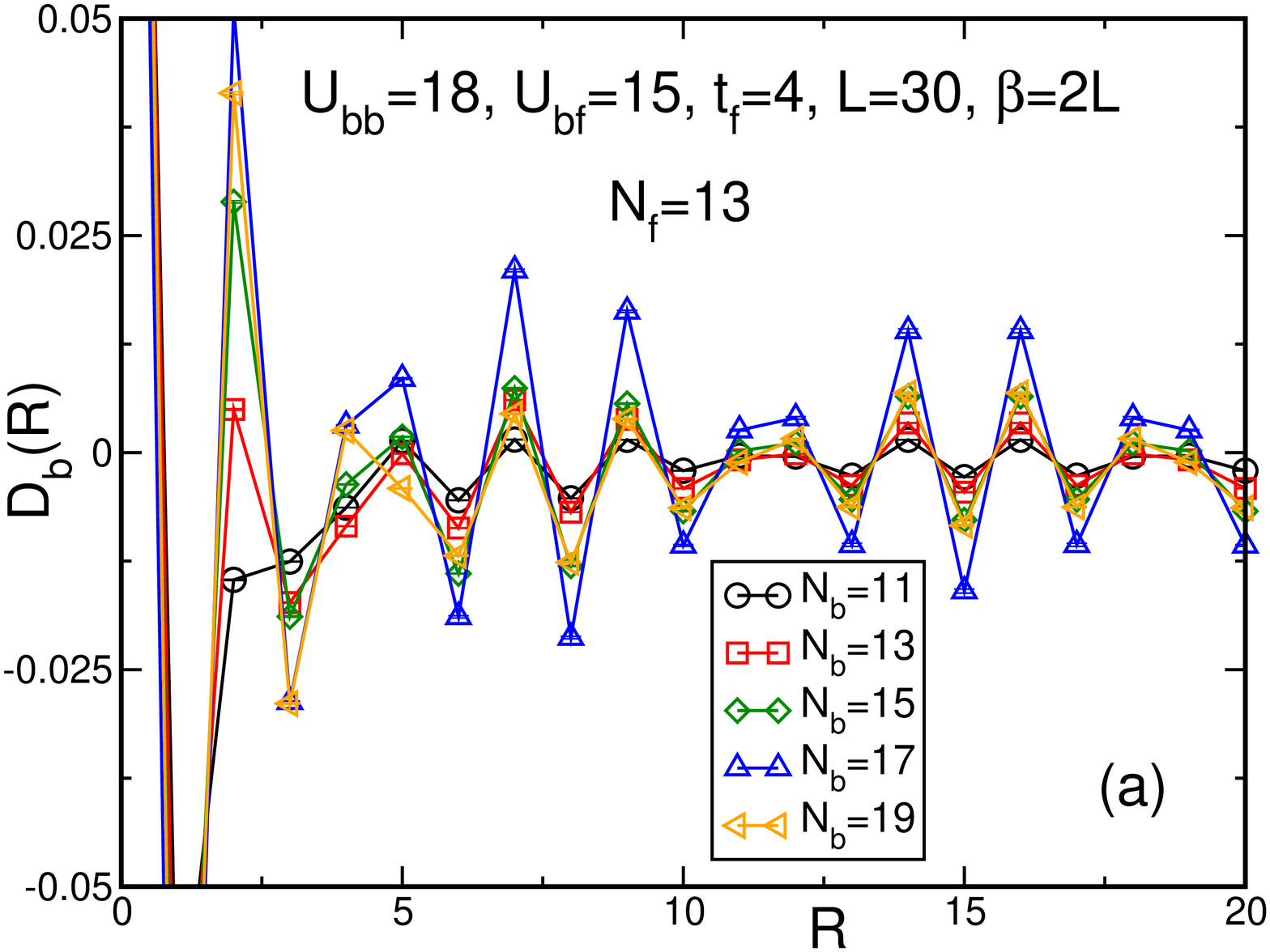}
\bigskip
\includegraphics[width=8.5cm]{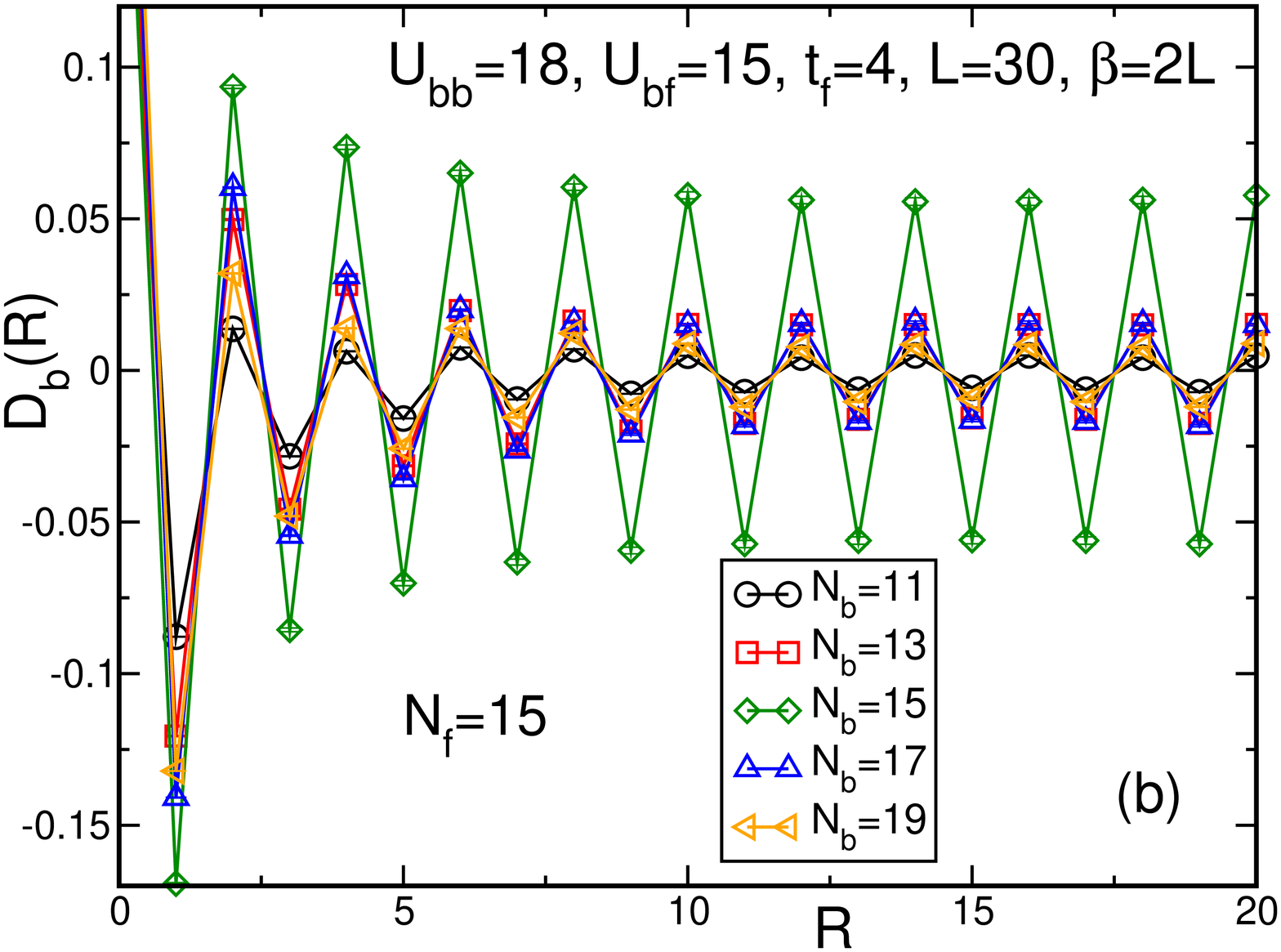}
\caption{(Color online) Density-density correlations for different
  boson fillings, $N_\text{b}$, and for two different fermion
  fillings, $N_\text{f} = L/2-2$ and $N_\text{f}=L/2$.  In the first
  case (a), we observe the beating characteristic of the pseudo spin
  phase at non zero magnetisation along $z$ with a maximum when
  $N_\text{f}+N_\text{b}=L$. In the second case (b), we see a long
  range order (or quasi long range) for the bosons.\label{doping}}
\end{figure}

Following this example, we consider a system where the repulsion
between bosons, $U_\text{bb} = 15$, is smaller than the boson-fermion
repulsion $U_\text{bf} =18$. Starting from the solid phase obtained
for $N_\text{b} = N_\text{f} = L / 2$ and $t_\text{f} = 4$,\cite{pollet06} we changed
slightly the number of the different types of particles and observed
how the solid order was modified (see Fig. \ref{doping}).

We found that upon changing the number of fermions, some defects are
introduced in the solid order, which are exposed by the characteristic
beating in the density-density correlations (Fig.~\ref{doping}, top).\cite{hebert05} 
On the other hand,
when one changes the number of bosons, the density wave order of
alternating empty and filled sites persists (Fig.~\ref{doping}, bottom). 
A surprising result is that
this wave order is present even when the number of bosons is reduced
below half filling, unlike what happens in the purely bosonic model
with near neighbor repulsion.\cite{batrouni06}

\begin{figure}[t]
\includegraphics[width=8.5cm]{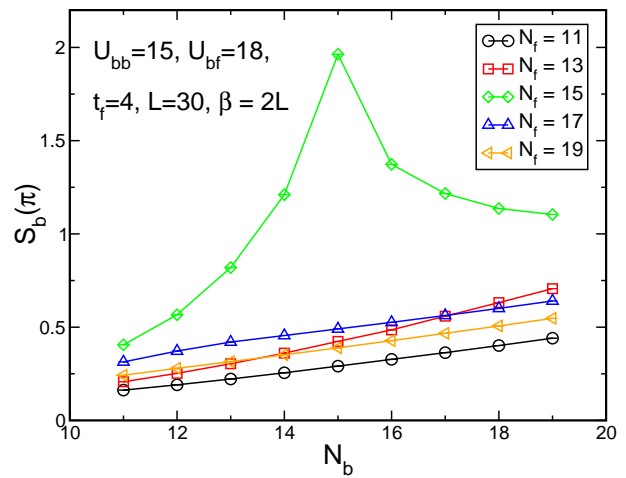}
\caption{(Color online) Structure factor $S_\text{b}(\pi)$ as a
  function of the number of bosons $N_\text{b}$ for different fermion
  fillings, $N_\text{f}$. The oscillations characterizing a density
  order only develop for $N_\text{f} = L/2$.\label{structurefactor}}
\end{figure}

Examining the bosonic structure factor $S_\text{b}(\pi)$
(Fig.~\ref{structurefactor}), we observe that only the case $N_f=L/2$
leads to large $S_\text{b}(\pi)$ and therefore to the long range density
order necessary for the establishment of SS. We also observe in
Fig.~\ref{structurefactor} that when the bosonic population is doped
above or below half filling, $S_\text{b}(\pi)$ drops but
remains rather appreciable especially above half filling. Finite size
scaling is required to establish if these non-vanishing values of
$S_\text{b}(\pi)$ persist in the thermodynamic limit (see section~\ref{sec:ffs}).

\begin{figure}[h]
\includegraphics[width=8.5cm]{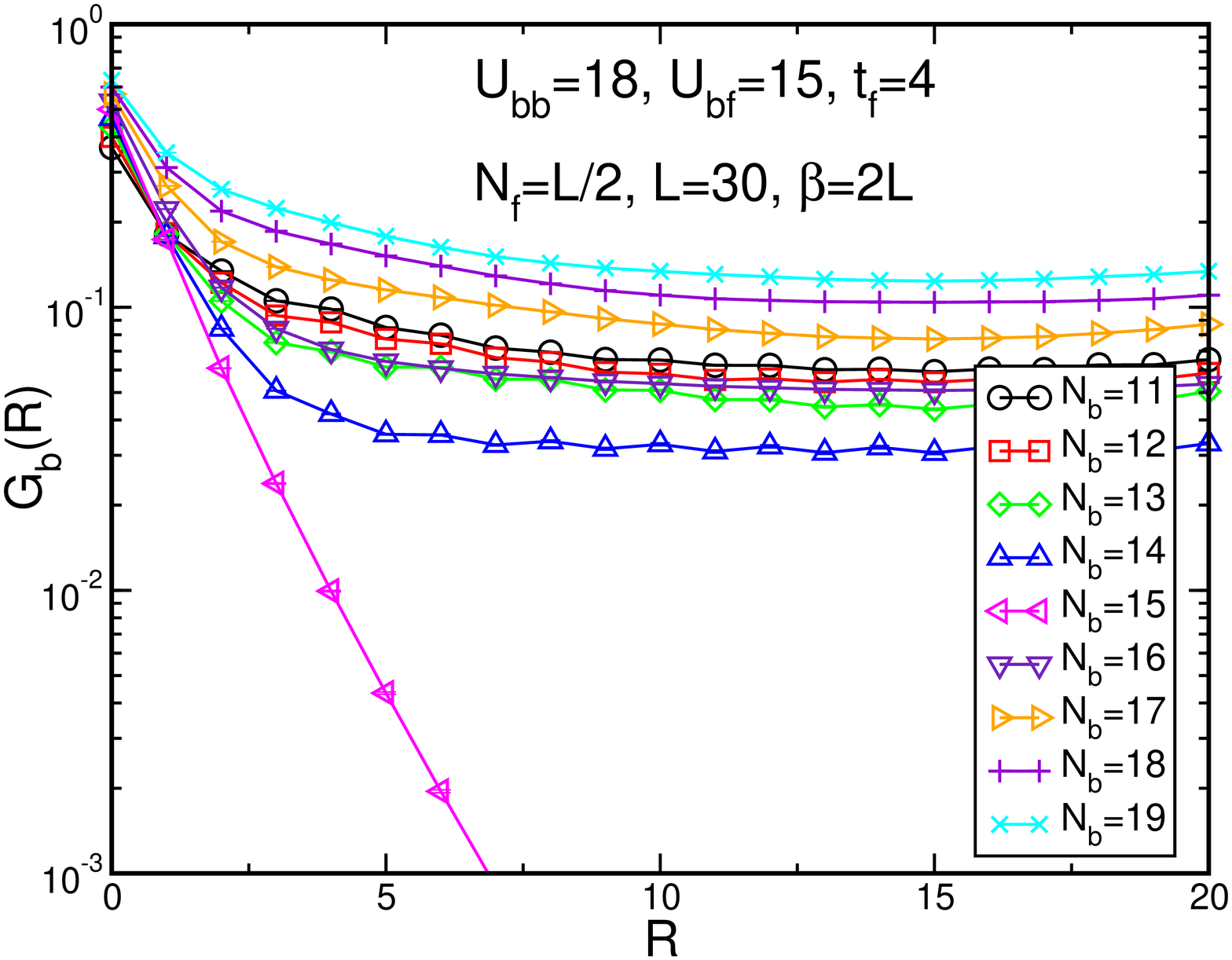}
\caption{(Color online) Phase correlation, $G_\text{b}(R)$, for
  different boson fillings.  There is always an algebraic decay
  characteristic of a superfluid in one dimension, except at double
  half-filling in the solid phase.\label{green}}
\end{figure}

As expected, quasi long range phase coherence is recovered as soon as
the solid is doped away from half-filling (see Fig.\ref{green}) and
the bosons become superfluid. This phase coherence is stronger when
the system is doped above half-filling.

These results lead to the conclusion that a supersolid phase can only
be found for $N_\text{f}=L/2$, where no defects are introduced in the
solid order. Doping the bosons above half-filling provides the best
candidate to observe a supersolid as both the structure factor and the
phase coherence are larger in this case but doping below half-filling
could also yield to a supersolid phase.  Finite size analysis of the
behavior of $\rho_\text{s}$ and $S_\text{b}(\pi)$ is needed to verify
if the supersolid phase persists in the thermodynamic limit.

\section{Finite size analysis \label{sec:ffs}}

In this system, the sizes that can be used in the finite size scaling
analysis are quite limited. In order to avoid sign problems,
$N_\text{f} = L/2$ must be odd. In addition, $L$ cannot be very large
in order for the simulation to converge in reasonable time.  We used
two different densities of bosons, $N_\text{b}/L = 0.4$ and 0.6 (one
below and one above half-filling) which give an integer number of
bosons for sizes $L = 10, 30, 50 \cdots$ When we could not obtain
exactly these densities with the given constraints, we used the number
of bosons that gives the closest density.

\begin{figure}[t]
\includegraphics[width=8.5cm]{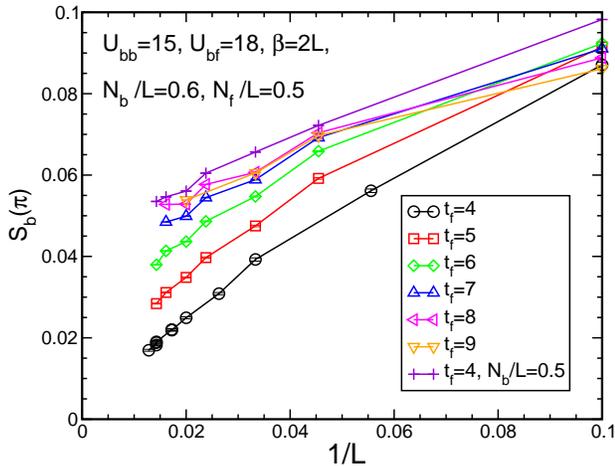}
\caption{(Color online) Finite size scaling of the structure factor
  $S_\text{b}(\pi)$ for different values of the hopping parameter
  $t_\text{f}$. The structure factor goes to zero in the large size
  limit for $t_\text{f}=4$.  For $t_\text{f} > 4$, in the range of
  sizes that are accessible, the structure factor is non
  zero.\label{FSS}}
\end{figure}

\begin{figure}[h]
\includegraphics[width=8.5cm]{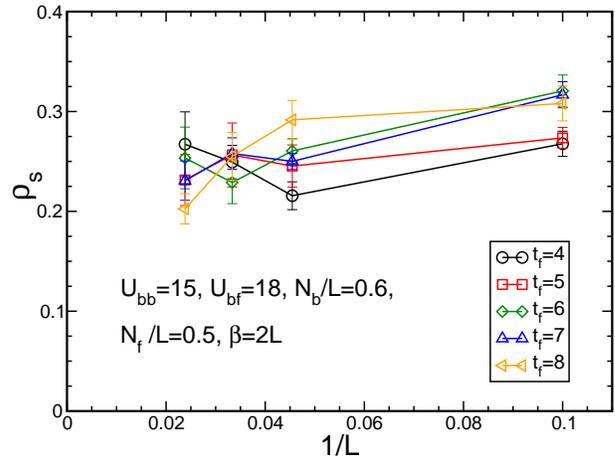}
\caption{(Color online) Finite size scaling of the superfluid density
$\rho_\text{s}$ in the case where the system is doped above
half-filling.  Reliable results for this quantity are difficult to
obtain for sizes larger than $L > 42$.  However, $\rho_s$ shows
very little variations when $L$ or $t_\text{f}$ vary and should then
remain non-zero in the large $L$ limit for all cases.
\label{SFdens}
}
\end{figure}

We first performed the finite size analysis with the parameters used
in the first part of this paper ($U_\text{bb}=18$, $U_\text{bf}=15$,
$t_\text{f}=4$) and above half-filling ($N_\text{b}/L=0.6$).  Figures
\ref{FSS} and \ref{SFdens} show that, for this case, the structure
factor goes to zero in the thermodynamic limit while the superfluid
density remains finite.  This indicates that what appears to be a
supersolid phase for small $L$ is in fact a superfluid.  Varying the
different available parameters ($U_\text{bb}$, $U_\text{bf}$,
$t_\text{f}$), we observed that increasing $t_\text{f}$ increases
noticeably the value of $S_\text{b}(\pi)$, whereas varying the
interaction did not yield similar variations.  For $t_\text{f}\geq 5$
(see Fig. \ref{FSS}) $S_\text{b}(\pi)$ extrapolates to a finite value
as $L$ increases. This indicates that for these values of $t_\text{f}$
the density wave order survives in the thermodynamic limit when
$N_\text{b} = 0.6 L$ and $N_\text{f} = L/2$.

\begin{figure}[b]
\includegraphics[width=8.5cm]{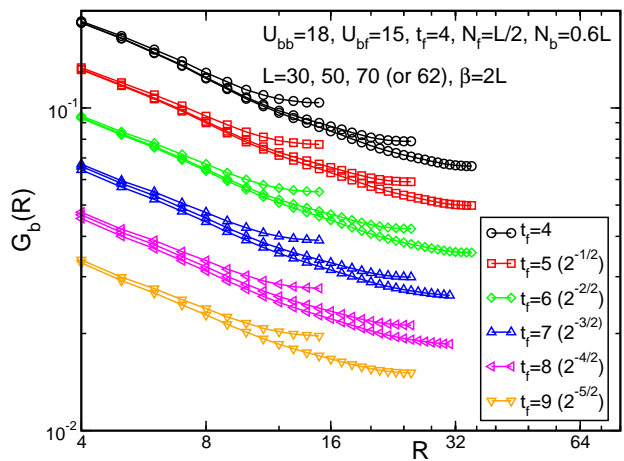}
\caption{(Color online) Finite size scaling of the Green function
  $G_\text{b}(R)$ for several values of the hopping parameter
  $t_\text{f}$.  $G_\text{b}(R)$ always appears to decay
  algebraically. The curves have been multiplied by a factor indicated
  in the legend between parentheses.\label{FSG}}
\end{figure}

In Fig.~\ref{SFdens} we see that the superfluid density changes very
little with $L$ or $t_\text{f}$: $\rho_s$ then goes to a finite value
when $L \rightarrow \infty$.  As mentionned above, obtaining precise
values for $\rho_s$ is difficult but, on the other hand, the Green
function, $G_\text{b}(R)$, is measured with very good statistical
accuracy and can, therefore, also be used to characterize the nature
of the phase coherence. Figure \ref{FSG} shows that $G_\text{b}(R)$
exhibits power law decay for all the sizes and values of the fermion
hopping parameter we have studied (for $N_\text{b} = 0.6 L$),
confirming the existence of a quasi condensate which leads to a
superfluid behavior in the presence of long range density order for
the fermions.

This indicates the presence of a supersolid phase in this system for
$t_\text{f}\geq 5$, $U_\text{bb}=18$, $U_\text{bf}=15$, and
$N_\text{b} = 0.6 L$.

\begin{figure}[t]
\includegraphics[width=8.5cm]{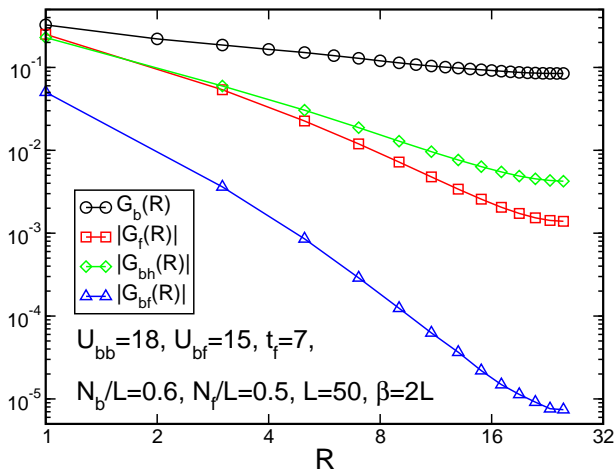}
\caption{(Color online) Green functions for bosons, fermions,
  boson-fermion pairs, and boson-hole pairs in the candidate
  supersolid phase. All the Green functions have algebraic
  decay.\label{otherG}}
\end{figure}

\begin{figure}[t]
\includegraphics[width=8.5cm]{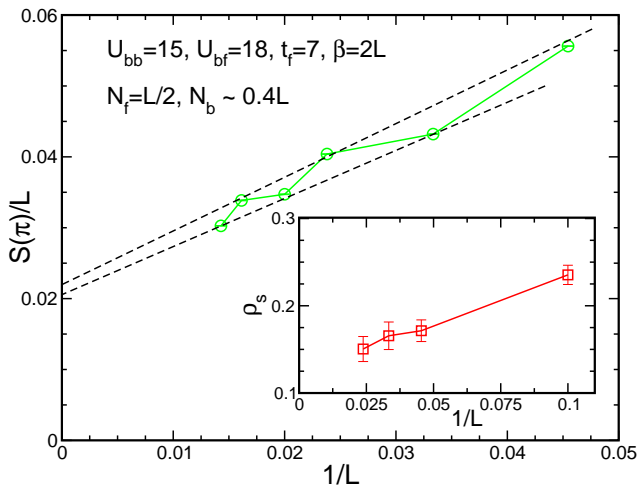}
\caption{(Color online) Finite size scaling of the structure factor
  $S_\text{b}(\pi)$ below half-filling ($N_\text{b}/L = 0.4$) for
  $t_\text{f} = 7$.  The structure factor goes to a non zero value in
  the large size limit.  The observed oscillations are due to the fact
  that we obtained exactly the desired density ($N_\text{b}/L = 0.4$)
  only for $L=30, 50, 70$.\label{belowHF}}
\end{figure}

So far, we have concentrated on the behavior of the bosons.  However,
the supersolid phase we have exposed is necessarily a collective phase
of fermions and bosons. To study the behavior of fermions or the
collective behavior of fermions and bosons, we introduce the Green
functions for fermions $G_\text{f}$, boson-fermion pairs
$G_\text{bf}$, and boson-hole pairs $G_\text{bh}$.

\begin{eqnarray}
G_\text{f}(R) &=& 
{1\over L} \sum_{r=1}^L 
\left\langle
f^\dagger_{r+R} f_r + f^\dagger_r f_{r+R}
\right\rangle\\
G_\text{bf}(R) &=& 
{1\over L} \sum_{r=1}^L 
\left\langle
f^\dagger_{r+R} b^\dagger_{r+R} b_r f_r + \text{h.c.}
\right\rangle\\
G_\text{bh}(R) &=& 
{1\over L} \sum_{r=1}^L 
\left\langle
f_{r+R} b^\dagger_{r+R} b_r f^\dagger_r + \text{h.c.} 
\right\rangle
\end{eqnarray}

Figure \ref{otherG} shows that these Green functions have algebraic
decay in the candidate supersolid phase. The dominant correlations are
those of individual bosons, they have the slowest decay and the
largest values. However, a description only in terms of bosons does
not fully characterize this phase; we see that individual fermion,
boson-hole and boson-fermion pairs also have algebraic phase
correlations and are also relevant degrees of freedom in the
system. The boson-hole pairs have much stronger phase correlations
than the boson-fermion pairs. This is to be expected since
anticorrelated movement of bosons and fermions is a generic phenomenon
in such mixtures with repulsive interactions
\cite{hebert07,kuklov03,pollet06}.

In a similar way to the case above half-filling, we performed finite
size scaling analysis when the system is doped below half-filling
($N_\text{b}/L = 0.4$). We studied a case where the value of $t_\text{f}$ is
quite large ($t_\text{f} = 7$) since we observed in the previous case
that it improves the structure factor. We find that, as for the case
above half filling, the structure factor and the superfluid density
both go to a finite value in the large $L$ limit and that a supersolid
phase is, therefore, thermodynamically stable (see
Fig. \ref{belowHF}). This behavior is different from what happens in
the bosonic case with near neighbor interactions where, doping below
half filling leads to a superfluid with soliton-like
quasiparticles.~\cite{batrouni06,hebert05} 

\section{conclusion}

In this paper we have studied, using exact QMC simulations, the
formation of supersolid phases in the ground state of Bose-Fermi
mixtures on optical lattices. At double half filling,
$N_\text{b}=N_\text{f}=L/2$, and for sufficiently large
$t_\text{f}/t_\text{b}$, the system exhibits long range density order
as exposed by $S_\text{b}(\pi)$.~\cite{pollet06} Inspired by the
behavior of the extended bosonic Hubbard model, which exhibits a
supersolid phase above half filling when the near neighbor repulsion
is large enough compared to the contact
term,~\cite{sengupta05,batrouni06}, we found that, similarly, when
$U_\text{bf} > U_\text{bb}$ and the system is doped by adding bosons,
the system enters a supersolid phase. It is important to keep in mind
that this phase is a collective Bose-Fermi phase: the various Green
functions (boson-boson, fermion-fermion and boson-fermion) we have
presented demonstrate this clearly.

Surprisingly, we have also found that upon doping the system below
half filling for the bosons, we also obtain a stable supersolid
phase. This is different from the behavior of the purely bosonic
system.

\begin{acknowledgments}
F. H. and G. G. B. are supported by the CNRS (France) PICS 18796 and
V. G. R. by the research program of the `Stichting voor Fundamenteel
Onderzoek der Materie (FOM)'. The authors would like to thank Lode
Pollet for providing us one of the program used in this study and for
many useful suggestions.
\end{acknowledgments}


\begin{thebibliography}{40}
\expandafter\ifx\csname natexlab\endcsname\relax\def\natexlab#1{#1}\fi
\expandafter\ifx\csname bibnamefont\endcsname\relax
  \def\bibnamefont#1{#1}\fi
\expandafter\ifx\csname bibfnamefont\endcsname\relax
  \def\bibfnamefont#1{#1}\fi
\expandafter\ifx\csname citenamefont\endcsname\relax
  \def\citenamefont#1{#1}\fi
\expandafter\ifx\csname url\endcsname\relax
  \def\url#1{\texttt{#1}}\fi
\expandafter\ifx\csname urlprefix\endcsname\relax\def\urlprefix{URL }\fi
\providecommand{\bibinfo}[2]{#2}
\providecommand{\eprint}[2][]{\url{#2}}

\bibitem[{\citenamefont{Penrose and Onsager}(1956)}]{penrose56}
\bibinfo{author}{\bibfnamefont{O.}~\bibnamefont{Penrose}} \bibnamefont{and}
  \bibinfo{author}{\bibfnamefont{L.}~\bibnamefont{Onsager}},
  \bibinfo{journal}{Phys. Rev.} \textbf{\bibinfo{volume}{104}},
  \bibinfo{pages}{576} (\bibinfo{year}{1956}).

\bibitem[{\citenamefont{\surname{Kim} and
  \surname{Chan}}(2004{\natexlab{a}})}]{kim04}
\bibinfo{author}{\bibfnamefont{E.}~\bibnamefont{\surname{Kim}}}
  \bibnamefont{and}
  \bibinfo{author}{\bibfnamefont{M.}~\bibnamefont{\surname{Chan}}},
  \bibinfo{journal}{Nature} \textbf{\bibinfo{volume}{427}},
  \bibinfo{pages}{225} (\bibinfo{year}{2004}{\natexlab{a}}).

\bibitem[{\citenamefont{\surname{Kim} and
  \surname{Chan}}(2004{\natexlab{b}})}]{kim04-2}
\bibinfo{author}{\bibfnamefont{E.}~\bibnamefont{\surname{Kim}}}
  \bibnamefont{and}
  \bibinfo{author}{\bibfnamefont{M.}~\bibnamefont{\surname{Chan}}},
  \bibinfo{journal}{Science} \textbf{\bibinfo{volume}{305}},
  \bibinfo{pages}{1941} (\bibinfo{year}{2004}{\natexlab{b}}).

\bibitem[{\citenamefont{\surname{Rittner} and
  \surname{Reppy}}(2006)}]{rittner06}
\bibinfo{author}{\bibfnamefont{A.}~\bibnamefont{\surname{Rittner}}}
  \bibnamefont{and}
  \bibinfo{author}{\bibfnamefont{J.}~\bibnamefont{\surname{Reppy}}},
  \bibinfo{journal}{Phys.\ Rev.\ Lett.} \textbf{\bibinfo{volume}{97}},
  \bibinfo{pages}{165301} (\bibinfo{year}{2006}).

\bibitem[{\citenamefont{\surname{Rittner} and
  \surname{Reppy}}(2007)}]{rittner07}
\bibinfo{author}{\bibfnamefont{A.}~\bibnamefont{\surname{Rittner}}}
  \bibnamefont{and}
  \bibinfo{author}{\bibfnamefont{J.}~\bibnamefont{\surname{Reppy}}},
  \bibinfo{journal}{Phys.\ Rev.\ Lett.} \textbf{\bibinfo{volume}{98}},
  \bibinfo{pages}{175302} (\bibinfo{year}{2007}).

\bibitem[{\citenamefont{Sasaki et~al.}(2006)\citenamefont{Sasaki, Ishiguro,
  Caupin, Maris, and Balibar}}]{sasaki06}
\bibinfo{author}{\bibfnamefont{S.}~\bibnamefont{Sasaki}},
  \bibinfo{author}{\bibfnamefont{R.}~\bibnamefont{Ishiguro}},
  \bibinfo{author}{\bibfnamefont{F.}~\bibnamefont{Caupin}},
  \bibinfo{author}{\bibfnamefont{H.}~\bibnamefont{Maris}}, \bibnamefont{and}
  \bibinfo{author}{\bibfnamefont{S.}~\bibnamefont{Balibar}},
  \bibinfo{journal}{Science} \textbf{\bibinfo{volume}{313}},
  \bibinfo{pages}{1098} (\bibinfo{year}{2006}).

\bibitem[{\citenamefont{\surname{Boninsegni}
  et~al.}(2006)\citenamefont{\surname{Boninsegni}, \surname{Kuklov},
  \surname{Pollet}, \surname{Prokof'ev}, \surname{Svistunov}, and
  \surname{Troyer}}}]{boninsegni06}
\bibinfo{author}{\bibfnamefont{M.}~\bibnamefont{\surname{Boninsegni}}},
  \bibinfo{author}{\bibfnamefont{A.}~\bibnamefont{\surname{Kuklov}}},
  \bibinfo{author}{\bibfnamefont{L.}~\bibnamefont{\surname{Pollet}}},
  \bibinfo{author}{\bibfnamefont{N.}~\bibnamefont{\surname{Prokof'ev}}},
  \bibinfo{author}{\bibfnamefont{B.}~\bibnamefont{\surname{Svistunov}}},
  \bibnamefont{and}
  \bibinfo{author}{\bibfnamefont{M.}~\bibnamefont{\surname{Troyer}}},
  \bibinfo{journal}{Phys.\ Rev.\ Lett.} \textbf{\bibinfo{volume}{97}},
  \bibinfo{pages}{080401} (\bibinfo{year}{2006}).

\bibitem[{\citenamefont{\surname{Pollet}
  et~al.}(2007)\citenamefont{\surname{Pollet}, \surname{Boninsegni},
  \surname{Kuklov}, \surname{Prokof'ev}, \surname{Svistunov}, and
  \surname{Troyer}}}]{boninsegni07}
\bibinfo{author}{\bibfnamefont{L.}~\bibnamefont{\surname{Pollet}}},
  \bibinfo{author}{\bibfnamefont{M.}~\bibnamefont{\surname{Boninsegni}}},
  \bibinfo{author}{\bibfnamefont{A.}~\bibnamefont{\surname{Kuklov}}},
  \bibinfo{author}{\bibfnamefont{N.}~\bibnamefont{\surname{Prokof'ev}}},
  \bibinfo{author}{\bibfnamefont{B.}~\bibnamefont{\surname{Svistunov}}},
  \bibnamefont{and}
  \bibinfo{author}{\bibfnamefont{M.}~\bibnamefont{\surname{Troyer}}},
  \bibinfo{journal}{Phys.\ Rev.\ Lett.} \textbf{\bibinfo{volume}{98}},
  \bibinfo{pages}{135301} (\bibinfo{year}{2007}).

\bibitem[{\citenamefont{\surname{Boninsegni}
  et~al.}(2007)\citenamefont{\surname{Boninsegni}, \surname{Kuklov},
  \surname{Pollet}, \surname{Prokof'ev}, \surname{Svistunov}, and
  \surname{Troyer}}}]{boninsegni07-2}
\bibinfo{author}{\bibfnamefont{M.}~\bibnamefont{\surname{Boninsegni}}},
  \bibinfo{author}{\bibfnamefont{A.}~\bibnamefont{\surname{Kuklov}}},
  \bibinfo{author}{\bibfnamefont{L.}~\bibnamefont{\surname{Pollet}}},
  \bibinfo{author}{\bibfnamefont{N.}~\bibnamefont{\surname{Prokof'ev}}},
  \bibinfo{author}{\bibfnamefont{B.}~\bibnamefont{\surname{Svistunov}}},
  \bibnamefont{and}
  \bibinfo{author}{\bibfnamefont{M.}~\bibnamefont{\surname{Troyer}}},
  \bibinfo{journal}{Phys.\ Rev.\ Lett.} \textbf{\bibinfo{volume}{99}},
  \bibinfo{pages}{035301} (\bibinfo{year}{2007}).

\bibitem[{\citenamefont{\surname{Batrouni}
  et~al.}(1990)\citenamefont{\surname{Batrouni}, \surname{Scalettar}, and
  \surname{Zimanyi}}}]{batrouni90}
\bibinfo{author}{\bibfnamefont{G.~G.} \bibnamefont{\surname{Batrouni}}},
  \bibinfo{author}{\bibfnamefont{R.~T.} \bibnamefont{\surname{Scalettar}}},
  \bibnamefont{and} \bibinfo{author}{\bibfnamefont{G.~T.}
  \bibnamefont{\surname{Zimanyi}}}, \bibinfo{journal}{Phys.\ Rev.\ Lett.}
  \textbf{\bibinfo{volume}{65}}, \bibinfo{pages}{1765} (\bibinfo{year}{1990}).

\bibitem[{\citenamefont{\surname{Batrouni} and Scalettar}(2000)}]{batrouni01}
\bibinfo{author}{\bibfnamefont{G.~G.} \bibnamefont{\surname{Batrouni}}}
  \bibnamefont{and} \bibinfo{author}{\bibfnamefont{R.~T.}
  \bibnamefont{Scalettar}}, \bibinfo{journal}{Phys.\ Rev.\ Lett.}
  \textbf{\bibinfo{volume}{84}}, \bibinfo{pages}{1599} (\bibinfo{year}{2000}).

\bibitem[{\citenamefont{Sengupta et~al.}(2005)\citenamefont{Sengupta, Pryadko,
  Alet, Troyer, and Schmid}}]{sengupta05}
\bibinfo{author}{\bibfnamefont{P.}~\bibnamefont{Sengupta}},
  \bibinfo{author}{\bibfnamefont{L.}~\bibnamefont{Pryadko}},
  \bibinfo{author}{\bibfnamefont{F.}~\bibnamefont{Alet}},
  \bibinfo{author}{\bibfnamefont{M.}~\bibnamefont{Troyer}}, \bibnamefont{and}
  \bibinfo{author}{\bibfnamefont{G.}~\bibnamefont{Schmid}},
  \bibinfo{journal}{Phys.\ Rev.\ Lett.} \textbf{\bibinfo{volume}{94}},
  \bibinfo{pages}{207202} (\bibinfo{year}{2005}).

\bibitem[{\citenamefont{Melko et~al.}(2005)\citenamefont{Melko, Paramekanti,
  Burkov, Vishwanath, Sheng, and Balents}}]{melko05}
\bibinfo{author}{\bibfnamefont{R.~G.} \bibnamefont{Melko}},
  \bibinfo{author}{\bibfnamefont{A.}~\bibnamefont{Paramekanti}},
  \bibinfo{author}{\bibfnamefont{A.~A.} \bibnamefont{Burkov}},
  \bibinfo{author}{\bibfnamefont{A.}~\bibnamefont{Vishwanath}},
  \bibinfo{author}{\bibfnamefont{D.~N.} \bibnamefont{Sheng}}, \bibnamefont{and}
  \bibinfo{author}{\bibfnamefont{L.}~\bibnamefont{Balents}},
  \bibinfo{journal}{Phys.\ Rev.\ Lett.} \textbf{\bibinfo{volume}{95}},
  \bibinfo{pages}{127207} (\bibinfo{year}{2005}).

\bibitem[{\citenamefont{Wessel and Troyer}(2005)}]{wessel05}
\bibinfo{author}{\bibfnamefont{S.}~\bibnamefont{Wessel}} \bibnamefont{and}
  \bibinfo{author}{\bibfnamefont{M.}~\bibnamefont{Troyer}},
  \bibinfo{journal}{Phys.\ Rev.\ Lett.} \textbf{\bibinfo{volume}{95}},
  \bibinfo{pages}{127205} (\bibinfo{year}{2005}).

\bibitem[{\citenamefont{Heidarian and Damle}(2005)}]{heidarian05}
\bibinfo{author}{\bibfnamefont{D.}~\bibnamefont{Heidarian}} \bibnamefont{and}
  \bibinfo{author}{\bibfnamefont{K.}~\bibnamefont{Damle}},
  \bibinfo{journal}{Phys.\ Rev.\ Lett.} \textbf{\bibinfo{volume}{95}},
  \bibinfo{pages}{127206} (\bibinfo{year}{2005}).

\bibitem[{\citenamefont{Boninsegni and Prokof'ev}(2005)}]{boninsegni05}
\bibinfo{author}{\bibfnamefont{M.}~\bibnamefont{Boninsegni}} \bibnamefont{and}
  \bibinfo{author}{\bibfnamefont{N.}~\bibnamefont{Prokof'ev}},
  \bibinfo{journal}{Phys.\ Rev.\ Lett.} \textbf{\bibinfo{volume}{95}},
  \bibinfo{pages}{237204} (\bibinfo{year}{2005}).

\bibitem[{\citenamefont{\surname{Batrouni}
  et~al.}(2006)\citenamefont{\surname{Batrouni}, \surname{H\'ebert}, and
  \surname{Scalettar}}}]{batrouni06}
\bibinfo{author}{\bibfnamefont{G.~G.} \bibnamefont{\surname{Batrouni}}},
  \bibinfo{author}{\bibfnamefont{F.}~\bibnamefont{\surname{H\'ebert}}},
  \bibnamefont{and} \bibinfo{author}{\bibfnamefont{R.~T.}
  \bibnamefont{\surname{Scalettar}}}, \bibinfo{journal}{Phys.\ Rev.\ Lett.}
  \textbf{\bibinfo{volume}{97}}, \bibinfo{pages}{087209}
  (\bibinfo{year}{2006}).

\bibitem[{\citenamefont{Dang et~al.}(2008)\citenamefont{Dang, Boninsegni, and
  Pollet}}]{pollet08}
\bibinfo{author}{\bibfnamefont{L.}~\bibnamefont{Dang}},
  \bibinfo{author}{\bibfnamefont{M.}~\bibnamefont{Boninsegni}},
  \bibnamefont{and} \bibinfo{author}{\bibfnamefont{L.}~\bibnamefont{Pollet}}
  (\bibinfo{year}{2008}), \eprint{arXiv:0803.1116}.

\bibitem[{\citenamefont{Chen et~al.}(2008)\citenamefont{Chen, Melko, Wessel,
  and Kao}}]{chen08}
\bibinfo{author}{\bibfnamefont{Y.-C.} \bibnamefont{Chen}},
  \bibinfo{author}{\bibfnamefont{R.~G.} \bibnamefont{Melko}},
  \bibinfo{author}{\bibfnamefont{S.}~\bibnamefont{Wessel}}, \bibnamefont{and}
  \bibinfo{author}{\bibfnamefont{Y.-J.} \bibnamefont{Kao}},
  \bibinfo{journal}{Phys.\ Rev.\ B} \textbf{\bibinfo{volume}{94}},
  \bibinfo{pages}{014524} (\bibinfo{year}{2008}).

\bibitem[{\citenamefont{Schreck et~al.}(2001)\citenamefont{Schreck, Khaykovich,
  Corwin, Ferrari, Bourdel, Cubizolles, and Salomon}}]{sympa1}
\bibinfo{author}{\bibfnamefont{F.}~\bibnamefont{Schreck}},
  \bibinfo{author}{\bibfnamefont{L.}~\bibnamefont{Khaykovich}},
  \bibinfo{author}{\bibfnamefont{K.~L.} \bibnamefont{Corwin}},
  \bibinfo{author}{\bibfnamefont{G.}~\bibnamefont{Ferrari}},
  \bibinfo{author}{\bibfnamefont{T.}~\bibnamefont{Bourdel}},
  \bibinfo{author}{\bibfnamefont{J.}~\bibnamefont{Cubizolles}},
  \bibnamefont{and} \bibinfo{author}{\bibfnamefont{C.}~\bibnamefont{Salomon}},
  \bibinfo{journal}{Phys. Rev. Lett.} \textbf{\bibinfo{volume}{87}},
  \bibinfo{pages}{080403} (\bibinfo{year}{2001}).

\bibitem[{\citenamefont{Hadzibabic et~al.}(2003)\citenamefont{Hadzibabic,
  Gupta, Stan, Schunck, Zwierlein, Dieckmann, and Ketterle}}]{sympa2}
\bibinfo{author}{\bibfnamefont{Z.}~\bibnamefont{Hadzibabic}},
  \bibinfo{author}{\bibfnamefont{S.}~\bibnamefont{Gupta}},
  \bibinfo{author}{\bibfnamefont{C.~A.} \bibnamefont{Stan}},
  \bibinfo{author}{\bibfnamefont{C.~H.} \bibnamefont{Schunck}},
  \bibinfo{author}{\bibfnamefont{M.~W.} \bibnamefont{Zwierlein}},
  \bibinfo{author}{\bibfnamefont{K.}~\bibnamefont{Dieckmann}},
  \bibnamefont{and} \bibinfo{author}{\bibfnamefont{W.}~\bibnamefont{Ketterle}},
  \bibinfo{journal}{Phys. Rev. Lett.} \textbf{\bibinfo{volume}{91}},
  \bibinfo{pages}{160401} (\bibinfo{year}{2003}).

\bibitem[{\citenamefont{Kuklov and Svistunov}(2003)}]{kuklov03}
\bibinfo{author}{\bibfnamefont{A.}~\bibnamefont{Kuklov}} \bibnamefont{and}
  \bibinfo{author}{\bibfnamefont{B.}~\bibnamefont{Svistunov}},
  \bibinfo{journal}{Phys.\ Rev.\ Lett.} \textbf{\bibinfo{volume}{90}},
  \bibinfo{pages}{100401} (\bibinfo{year}{2003}).

\bibitem[{\citenamefont{Duan et~al.}(2003)\citenamefont{Duan, Demler, and
  Lukin}}]{duan03}
\bibinfo{author}{\bibfnamefont{L.-M.} \bibnamefont{Duan}},
  \bibinfo{author}{\bibfnamefont{E.}~\bibnamefont{Demler}}, \bibnamefont{and}
  \bibinfo{author}{\bibfnamefont{M.~D.} \bibnamefont{Lukin}},
  \bibinfo{journal}{Phys. Rev. Lett.} \textbf{\bibinfo{volume}{91}},
  \bibinfo{pages}{090402} (\bibinfo{year}{2003}).

\bibitem[{\citenamefont{Sengupta and Pryadko}(2005)}]{sengupta05-2}
\bibinfo{author}{\bibfnamefont{P.}~\bibnamefont{Sengupta}} \bibnamefont{and}
  \bibinfo{author}{\bibfnamefont{L.}~\bibnamefont{Pryadko}}
  (\bibinfo{year}{2005}), \eprint{cond-mat/0512241}.

\bibitem[{\citenamefont{Pollet et~al.}(2006)\citenamefont{Pollet, Troyer,
  Houcke, and Rombouts}}]{pollet06}
\bibinfo{author}{\bibfnamefont{L.}~\bibnamefont{Pollet}},
  \bibinfo{author}{\bibfnamefont{M.}~\bibnamefont{Troyer}},
  \bibinfo{author}{\bibfnamefont{K.~V.} \bibnamefont{Houcke}},
  \bibnamefont{and} \bibinfo{author}{\bibfnamefont{S.~M.~A.}
  \bibnamefont{Rombouts}}, \bibinfo{journal}{Phys. Rev. Lett.}
  \textbf{\bibinfo{volume}{96}}, \bibinfo{pages}{190402}
  (\bibinfo{year}{2006}).

\bibitem[{\citenamefont{Mathey et~al.}(2004)\citenamefont{Mathey, Wang,
  Hofstetter, Lukin, and Demler}}]{mathey04}
\bibinfo{author}{\bibfnamefont{L.}~\bibnamefont{Mathey}},
  \bibinfo{author}{\bibfnamefont{D.-W.} \bibnamefont{Wang}},
  \bibinfo{author}{\bibfnamefont{W.}~\bibnamefont{Hofstetter}},
  \bibinfo{author}{\bibfnamefont{M.~D.} \bibnamefont{Lukin}}, \bibnamefont{and}
  \bibinfo{author}{\bibfnamefont{E.}~\bibnamefont{Demler}},
  \bibinfo{journal}{Phys. Rev. Lett.} \textbf{\bibinfo{volume}{93}},
  \bibinfo{pages}{120404} (\bibinfo{year}{2004}).

\bibitem[{\citenamefont{Mathey and Wang}(2007)}]{mathey07}
\bibinfo{author}{\bibfnamefont{L.}~\bibnamefont{Mathey}} \bibnamefont{and}
  \bibinfo{author}{\bibfnamefont{D.-W.} \bibnamefont{Wang}},
  \bibinfo{journal}{Phys. Rev. A} \textbf{\bibinfo{volume}{75}},
  \bibinfo{eid}{013612} (\bibinfo{year}{2007}).

\bibitem[{\citenamefont{Titvinidze et~al.}(2008)\citenamefont{Titvinidze,
  Snoek, and Hofstetter}}]{titvinidze08}
\bibinfo{author}{\bibfnamefont{I.}~\bibnamefont{Titvinidze}},
  \bibinfo{author}{\bibfnamefont{M.}~\bibnamefont{Snoek}}, \bibnamefont{and}
  \bibinfo{author}{\bibfnamefont{W.}~\bibnamefont{Hofstetter}},
  \bibinfo{journal}{Phys. Rev. Lett.} \textbf{\bibinfo{volume}{100}},
  \bibinfo{eid}{100401} (\bibinfo{year}{2008}).

\bibitem[{\citenamefont{B\"uchler and Blatter}(2003)}]{buchler03}
\bibinfo{author}{\bibfnamefont{H.~P.} \bibnamefont{B\"uchler}}
  \bibnamefont{and} \bibinfo{author}{\bibfnamefont{G.}~\bibnamefont{Blatter}},
  \bibinfo{journal}{Phys. Rev. Lett.} \textbf{\bibinfo{volume}{91}},
  \bibinfo{pages}{130404} (\bibinfo{year}{2003}).

\bibitem[{\citenamefont{Mathey et~al.}(2008)\citenamefont{Mathey, Danshita, and
  Clark}}]{mathey08}
\bibinfo{author}{\bibfnamefont{L.}~\bibnamefont{Mathey}},
  \bibinfo{author}{\bibfnamefont{I.}~\bibnamefont{Danshita}}, \bibnamefont{and}
  \bibinfo{author}{\bibfnamefont{C.~W.} \bibnamefont{Clark}}
  (\bibinfo{year}{2008}), \eprint{arXiv:0806.0461}.

\bibitem[{\citenamefont{Mathey}(2007)}]{mathey07-2}
\bibinfo{author}{\bibfnamefont{L.}~\bibnamefont{Mathey}},
  \bibinfo{journal}{Phys. Rev. B} \textbf{\bibinfo{volume}{75}},
  \bibinfo{pages}{144510} (\bibinfo{year}{2007}).

\bibitem[{\citenamefont{Pollock and Ceperley}(1987)}]{pollock87}
\bibinfo{author}{\bibfnamefont{E.~L.} \bibnamefont{Pollock}} \bibnamefont{and}
  \bibinfo{author}{\bibfnamefont{D.~M.} \bibnamefont{Ceperley}},
  \bibinfo{journal}{Phys. Rev. B} \textbf{\bibinfo{volume}{36}},
  \bibinfo{pages}{8343} (\bibinfo{year}{1987}).

\bibitem[{\citenamefont{Rombouts et~al.}(2006)\citenamefont{Rombouts, Houcke,
  and Pollet}}]{rombouts06}
\bibinfo{author}{\bibfnamefont{S.~M.~A.} \bibnamefont{Rombouts}},
  \bibinfo{author}{\bibfnamefont{K.~V.} \bibnamefont{Houcke}},
  \bibnamefont{and} \bibinfo{author}{\bibfnamefont{L.}~\bibnamefont{Pollet}},
  \bibinfo{journal}{Phys.\ Rev.\ Lett.} \textbf{\bibinfo{volume}{96}},
  \bibinfo{pages}{180603} (\bibinfo{year}{2006}).

\bibitem[{\citenamefont{\surname{Van Houcke}
  et~al.}(2006)\citenamefont{\surname{Van Houcke}, Rombouts, and
  Pollet}}]{vanhoucke06}
\bibinfo{author}{\bibfnamefont{K.}~\bibnamefont{\surname{Van Houcke}}},
  \bibinfo{author}{\bibfnamefont{S.~M.~A.} \bibnamefont{Rombouts}},
  \bibnamefont{and} \bibinfo{author}{\bibfnamefont{L.}~\bibnamefont{Pollet}},
  \bibinfo{journal}{Phys.\ Rev.\ E} \textbf{\bibinfo{volume}{73}},
  \bibinfo{pages}{056703} (\bibinfo{year}{2006}).

\bibitem[{\citenamefont{Rousseau}(2008{\natexlab{a}})}]{rousseau08}
\bibinfo{author}{\bibfnamefont{V.~G.} \bibnamefont{Rousseau}},
  \bibinfo{journal}{Phys. Rev. E} \textbf{\bibinfo{volume}{77}},
  \bibinfo{pages}{056705} (\bibinfo{year}{2008}{\natexlab{a}}).

\bibitem[{\citenamefont{Rousseau}(2008{\natexlab{b}})}]{rousseau08-2}
\bibinfo{author}{\bibfnamefont{V.~G.} \bibnamefont{Rousseau}}
  (\bibinfo{year}{2008}{\natexlab{b}}), \eprint{arXiv:0806.1410}.

\bibitem[{\citenamefont{Pollet}(2005)}]{pollet05}
\bibinfo{author}{\bibfnamefont{L.}~\bibnamefont{Pollet}}, Ph.D. thesis,
  \bibinfo{school}{Universiteit Gent} (\bibinfo{year}{2005}),
  \eprint{http://www.nustruc.ugent.be/phdtheses.htm}.

\bibitem[{\citenamefont{Giamarchi}(2003)}]{giam03}
\bibinfo{author}{\bibfnamefont{T.}~\bibnamefont{Giamarchi}},
  \emph{\bibinfo{title}{Quantum Physics in One Dimension}}
  (\bibinfo{publisher}{Oxford University Press}, \bibinfo{year}{2003}).

\bibitem[{\citenamefont{H\'ebert et~al.}(2005)\citenamefont{H\'ebert,
  \surname{Batrouni}, and \surname{Scalettar}}}]{hebert05}
\bibinfo{author}{\bibfnamefont{F.}~\bibnamefont{H\'ebert}},
  \bibinfo{author}{\bibfnamefont{G.~G.} \bibnamefont{\surname{Batrouni}}},
  \bibnamefont{and} \bibinfo{author}{\bibfnamefont{R.~T.}
  \bibnamefont{\surname{Scalettar}}}, \bibinfo{journal}{Phys.\ Rev.\ A}
  \textbf{\bibinfo{volume}{71}}, \bibinfo{pages}{063609}
  (\bibinfo{year}{2005}).

\bibitem[{\citenamefont{H\'{e}bert et~al.}(2007)\citenamefont{H\'{e}bert,
  Haudin, Pollet, and Batrouni}}]{hebert07}
\bibinfo{author}{\bibfnamefont{F.}~\bibnamefont{H\'{e}bert}},
  \bibinfo{author}{\bibfnamefont{F.}~\bibnamefont{Haudin}},
  \bibinfo{author}{\bibfnamefont{L.}~\bibnamefont{Pollet}}, \bibnamefont{and}
  \bibinfo{author}{\bibfnamefont{G.~G.} \bibnamefont{Batrouni}},
  \bibinfo{journal}{Phys. Rev. A} \textbf{\bibinfo{volume}{76}},
  \bibinfo{eid}{043619} (\bibinfo{year}{2007}).

\end{thebibliography}

\end{document}